\documentclass[12pt]{article}

\usepackage{amsmath,amsthm,amssymb,color,undertilde}
\usepackage{epsfig,amsfonts,latexsym, hyperref, bbm, mathrsfs}
\usepackage{graphicx}
\makeatletter
\newcommand*\bigcdot{\mathpalette\bigcdot@{.5}}
\newcommand*\bigcdot@[2]{\mathbin{\vcenter{\hbox{\scalebox{#2}{$\m@th#1\bullet$}}}}}
\makeatother
\begin{document}	
\begin{center}
	{\bf\large{Geometric Phases for Classical and Quantum  Dynamics:\\ Hannay angle and Berry Phase for Loops on a Torus}}\\
	\vspace {3cm}
	{\bf{Subir Ghosh}}\\
	
	\vspace {1.5cm}
	Physics and Applied Mathematics Unit,\\
Indian Statistical Institute,\\
203 B. T. Road, Kolkata 700108, India.
		
\end{center}
	
	\vspace{1cm}
	{\bf{Abstract:}} 	In this paper we have considered closed trajectories of a particle on a two-torus where the loops are   noncontractible (poloidal and toroidal loops and knots embedded on a regular torus).    We have  calculated Hannay angle and Berry phase for particle traversing such  loops and knots when the torus itself is adiabatically  revolving. Since noncontractible loops do not enclose any area  Stokes theorem has to be applied with caution. In our computational scheme we have worked with line integrals directly thus avoiding Stokes theorem.    
	
	\vskip 1cm
{\bf{(1) Introduction:}}\\
Classical limit to quantum phenomena is much more than a naive $\hbar\rightarrow 0$. In this context the relation between two geometrical objects, Berry phase \cite{b1} in quantum regime and Hannay angle \cite{h1} in classical domain is indeed fascinating and has generated a lot of interest in theoretical and experimental physics. Quite interestingly, contrary to common occurrences, Berry phase (the quantum effect) was discovered first which was followed by its classical counterpart, Hannay angle and the delicate issue of identifying the latter as a semi-classical limit of the former was rigorously settled by Berry \cite{b3}. As a quantum system traverses a closed loop in   parameter space the anholonomy, (in Hilbert space where the system lives), is responsible of the appearance of Berry phase. On the other hand, Hannay angle is generated by anholonomy in the trajectory of the classical system in physical configuration space (or  more generally in the abstract action-angle manifold).  

An important generalization to multi-dimensional systems, (consisting of more than one degrees of freedom), where the quantum system in  classical limit  can show chaotic behavior, was performed by Robbins in \cite{rob} (hereafter referred to as JR). Precursors to this work are \cite{rob1} where physical effects of (the classical limit of) geometric $2$-form manifests in a Lorentz force form of "geometric magnetic" force {\footnote{This turned out to be the anti-symmetric partner of "deterministic friction", a dissipative force proposed by Wilkinson \cite{wil}.}}. However, an interesting question, as pointed out in \cite{b2}, was whether the classical limit of geometric phase can appear directly from adiabatic cyclic transformations of classical chaotic systems.  Early ideas that proposed the importance of geometric phases were the Pancharatnam phase and Aharanov-Bohm phase \cite{panch}. An exhaustive list of theoretical and experimental works are found in \cite{coll}. We list in \cite{berrnew} some recent applications of Berry Phase in areas related to quantum computation. Applications of Hannay angle in diverse fields (both theoretical and experimental) are mentioned in \cite{colhan}.

The present work deals with explicit evaluation and comparison of Hannay angle and Berry phase for the classical problem of a particle moving in closed trajectories on the surface of a two-torus and in its quantum version respectively. The problem is a non-trivial extension of the conventional one: geometric phase for particle on a sphere. The difference lies in the fact that on the sphere, a genus zero manifold, all closed non-intersecting loops are contractible whereas the torus, a genus one manifold, admits non-contractible loops of two varieties, toroidal and poloidal (to be explained later) as well as loops with knots.

In this context it will be worthwhile to discuss a little more in detail the basic results of JR \cite{rob} since our work can be thought of as a special case of JR where the system is reduced to a single  degree of freedom.  The  $n$-degrees of freedom classical  Hamiltonian systems in JR  is ergodic on the $(2n-1)$-dimensional energy shell in phase space {\footnote{An informal  way to define ergodicity is that the system finishes by being more or less everywhere in phase space if one waits long enough.}}. A particle restricted to move on a loop constitute a single  degree of freedom  special case where the (connected) energy shell coincides with the classical orbit.
 Furthermore, the framework of JR has the added advantage that  (adiabatic) variations in the Hamiltonian are interpreted as unitary and canonical transformations in the quantum and classical cases respectively. A special case that is of relevance is adiabatic rotations for which the generators are the components of angular rotation.  A key feature for a many particle sytem is to introduce a micro-canonical averaged variable, eg. angular momentum $<{\textbf{L}}>_E$ {\footnote{The definition of a generic $f$ is given in \cite{rob}, $<f>_E=(1/\partial_E\Omega )\int~dz~\delta(E-h)f(z,\textbf{R})$ where the normalization is the phase space volume on the energy shell $\partial_E\Omega =\int~dz~\delta(E-h)$ satisfying $h(z,{\textbf R})=E$. The total phase space volume inside the $E$-shell $\Omega (E)=\int~dz~\delta(E-h)$ is independent of $\textbf{R}$.}} for a given energy $E$ that amounts to $<{\textbf{L}}>_E=2v{\textbf{A}}/C$ where where $v$ is the (constant) velocity, $C$ is the length of the loop and ${\textbf{A}} =(A_1,A_2,A_3)$ is the vector of projected areas of the loop onto the coordinate planes.

 The results of JR apply to systems for which the  adiabatic rotations are a particular case; it is shown  that the classical limit of the Berry one-form applied to angular velocity   $\textbf w $ (regarded as a tangent vector in the parameter space of rotations) is given by $-<{\textbf{L}}>_E\bigcdot\,\textbf w$, where $<{\textbf{L}}>_E$ is the microcanonical average of the  angular momentum, and $E$ corresponds to the quantum number $n$ via the  semiclassical quantization rule. The Hannay one-form is then shown to be
 the derivative of the classical limit of the Berry one-form with respect to the
 classical action. Expressions for the Berry and Hannay two-forms are also
 given in JR.  A free particle constrained to move on a closed loop is an example of such a
 system.  Since the Berry one-form is linear in the classical action per unit mass $vC/2$,
 one gets that the Hannay 1-form applied to angular velocity $\hat{\textbf w}$ as $-4\pi {\textbf{A}}\bigcdot\,\hat{\textbf w} /C^2 $. Under an adiabatic $2\pi$-rotation with fixed angular velocity $\bf w$, the
 Hannay angle is given by $   −8\pi^2{\textbf{A}}\bigcdot\,\hat{\textbf w} /C^2$, 
  where $\hat{\textbf w}$ is the unit vector along the  axis of rotation. Thus  Hannay angle corresponds to a displacement along the loop of length
    \begin{equation}
   d=-4\pi {\textbf{A}}\bigcdot\,\hat{\textbf w} /C .
   \label{hann}
    \end{equation}
    Let us note that this result corroborates nicely with a general statement by Hannay \cite{h2} as follows, "(Hannay) angle, or rather the associated shift, for a general rigid rotation of a general loop (planar or not) in space about a fixed axis is ($4 \pi$ times) the projected area of the loop onto a plane perpendicular to the rotation axis divided by the length of the loop". This is also true for the Berry phase in the limit that the quantum particle is strictly confined
 to the loop.

 Let us now clarify the perspective of our analysis. Conventionally Berry phase or Hannay angle are expressed in terms of area (or solid angle) associated with the closed path since {\it{in generic examples the closed loops or  cycles    are taken to be bounding cycles}} meaning that they enclose an area in ${\bf{R}}$ (or in a surface embedded in ${\bf{R}}$).  Subsequently Stokes law is applied to introduce the area enclosed by the loop to finally reproduce the well known results in question, {\it{i.e.}} anholonomy, Berry Phase or Hannay angle. In standard examples one considers the parameter space to be  $S^2$ in ${\bf{R}}$ on which all closed loops are bounding. We will always talk about non-self-intersecting loops{\footnote{I thank Professor Berry for pointing this out.}}. 
 
 It appears that a straightforward application of Stokes theorem to convert line integrals to surface integrals (to get Hannay angle and Berry phase in the well known forms) is problematic since the non-contractible loops on a torus do not enclose any area. For the anholonomy on a torus it is possible to use the following prescription \cite{subir}: To define enclosed area by  noncontractable loops (in order to apply Stokes theorem) a generalization has been suggested by Hannay \cite{h2} where one has to consider a reference loop in conjunction with the loop in question and join them by a thin neck. A comment about the reference loop is in order. We point out that  on a compact two-dimensional non-simply-connected Riemannian
 manifold, each homotopy class of closed loops contains at least one
 geodesic, for which the holonomy necessarily vanishes. Hence it is natural to choose the geodesic as the reference loop. In the line integral the contribution of the neck cancels out and one is left with only the contribution of the loop in study. The advantage is that this extended loop structure encloses an area in the conventional way so that Stokes theorem can, in principle, be applied for closed paths with knots (and unknots). We, on the other hand, have computed the translational anholonomy following the line integral approach in \cite{subir}.

	  Surfaces having "handles" allows such possibility of noncontractible loops, torus being the simplest such example of a surface. Toroidal (fixed $\theta$)  and poloidal (fixed $\phi$)  loops  (see Figure 1) or a combination of both in the form of knot or unknot  are examples of noncontractible cycles \cite{knot}. In this paper we will consider  the dynamical problem of a particle  traversing these cycles on a slowly {\it{revolving}} torus and compute  the associated   Hannay angle and Berry phase in classical and quantal scenarios respectively.  Very recently classical and quantum dynamics of particle on torus has generated a lot of interest \cite{sree}. (Earlier important works where particle dynamics on torus is relevant are cited in \cite{sree}.) In this context it needs to be pointed out that even though the specific trajectories are noncontractible still what matters is the contractibility and anholonomy (or its absence) of the {\it{parameter}} space, which in the presented examples is still $S^1$, ie. space of rotations about a fixed axis. Hence, following the formalism mentioned below,  it is still possible to determine the Hannay angle the surface spanned by a specific surface. 	
	   For
	 the particle constrained to move on a closed loop in $R3$, one may consider the parameter space to be $SO(3)$, the space of all rotations.
	 Note that all cycles of $2\pi$-rotations about a fixed axis, regarded as closed
	 curves in $SO(3)$, are homotopic to each other. The result (\ref{hann}) shows that Hannay angle vanishes for  rotation about an axis $\textbf {p}$ perpendicular to $\textbf {A}$. Then it is possible to reproduce 
	 Hannay angle for a complete rotation about an arbitrary axis $\textbf {n}$ by the
	 integral of the curvature over any surface spanned  between rotations about $\textbf {p}$ and $\textbf {n}$. Incidentally for an explicit example having vanishing curvature but non-vanishing holonomy (see \cite{monod}).
	
	According to correspondence principle the spectral invariants of a quantum system are related  to invariant manifolds of classical system. For classical ergodicity the motion is confined to $2n-1$-dimensional constant energy shell and for integrable systems these are reduced to invariant tori. For our system consisting of a single particle moving on a loop, (for connected energy shell), the constant energy manifold gets identified with the particle trajectory. 
	
	Before going to the nitty-gritty of the present work let us emphasize that the present, apparently artificial looking model of a point particle sliding on a revolving torus, is not of purely academic interest. in fact it has similarities with a general area of interest, {\it i.e.} time dependent  Aharonov-Bohm effect (see for example \cite{ah1,ah01,ah2}).  As pointed out in \cite{ah1} explicit time dependence in a multiply connected coordinate space induces some amount of ambiguity in defining physical variables. In fact \cite{ah2} specifically studies Stokes theorem in similar context and reveals a surprising result that Stokes theorem can be satisfied only in a particular gauge. These studies are important in the context of recent practical applications \cite{ah3}.

	 The paper is organized as follows. In Section 2 Hannay angle \cite{h1} is computed for  particle dynamics on a revolving torus loop from two different frameworks \cite{b3}).  Then in Section 3, we will derive the Berry phase \cite{b1} for the corresponding quantal cases and  comment upon  the formula derived by Berry that connects the two geometric objects \cite{b3}.
	 In Section 4 we will conclude  with mentioning directions for further study.
	\vskip .3cm
	
{\bf{(2) Noncontractible loops and Hannay angle}}

In \cite{h1} Hannay has studied motion of a particle sliding frictionlessly on a planar hoop of arbitrary shape. Simultaneously  the hoop itself is also revolving about an axis perpendicular to the plane of the hoop. Hannay angle is the difference between  positions of the particle on the hoop, with and without revolution of the hoop itself. This geometric object depends only on the area enclosed by the hoop. Stokes theorem can be used (although it is not essential) to show the result for an irregular  hoop. In the present work we extend the analysis to  a particle  moving on a regular torus with the torus itself revolving in space. The hoop has a first homotopy group $\pi_1 (hoop)\approx Z^1$ whereas for the  torus it is  $\pi_1 (torus)\approx Z^1\times Z^1$ hence the extension  is non-trivial due to the   nature of noncontractable loops. 

\begin{figure}[htb!]
	{\centerline{\includegraphics[width=6cm, height=6cm] {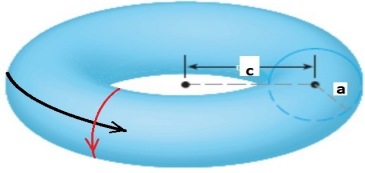}}}
	\caption{Red and black directions refer to poloidal ($\theta $) and toroidal ($\phi $) directions. Parameters $c$ and $a$ for a toroidal surface are shown.} \label{fig1}
\end{figure}

The generic relation $d/dt \mid_{space} =d/dt \mid_{body}+{\bf{\Omega}}\times $ \cite{h1}  connects time derivatives of a vector between  static frame  and a frame rotating with ${\bf{\Omega }}$. Applied to the position ${\bf{r}}$ of a unit mass  particle we find
$
\bf p\mid_{space} =\bf p \mid_{body}+\bf\Omega \times \bf r$, where  momenta of a unit mass particle are related. Action, {\it{i.e.}} the line integral around the loop, is an adiabetic invariant given by, 
\begin{equation}
\label{h2}
I=\frac{1}{2\pi}\oint \bf p \bigcdot\, d\bf r =\frac{1}{2\pi}\oint (\dot{\bf r}+\bf\Omega \times \bf r)\bigcdot\, d\bf r \,. 
\end{equation}
If  Stokes  theorem is applicable, the second term is expressed  as $2{\bf{\Omega }}\bigcdot \,{\bf{A}}/L $ where $A$ and $L$ are the area enclosed by the loop and length of the loop respectively. This term leads to the shift in angle for the particle motion  $\Delta \theta =-8\pi^2A/L^2 $ that is the Hannay angle \cite{h1}. However in the present work we will use the  line integral prescription since we focus on noncontractible loops.

In an alternative way   the Hannay angle was revealed by  Berry \cite{b3} by considering Newtonian dynamics with Euler pseudo forces that operate in the non-inertial rotating (torus) frame. It can be shown \cite{b3}
\begin{equation}
\label{h3}
\ddot{s}(t)=\bf t \bigcdot\, (-\bf\Omega \times (\bf\Omega\times \bf r)-\dot{\bf\Omega }\times r )=\frac{d{\bf r} (s(t))}{ds}\bigcdot\,(\Omega^2{\bf r}-({\bf \Omega}.{\bf r}){\bf \Omega}-\dot{\bf\Omega }\times r )
\end{equation}
where $s(t)$ is an arc length measured from a specific point on the loop and $\bf{t}$ is the unit tangent vector. The Coreolis force $\sim~\bf\Omega \times \dot{\bf r}$ does not appear since the particle is restricted to move in a fixed path and this force is normal to the path. Two integrations lead to the implicit solution,
\begin{equation}
\label{bb1}
s(t)=s_0+p_0t +\int_0^t dt' (t-t')F(t',s)
\end{equation}
where we have clubbed together all the terms of RHS of (\ref{h3}) as $F(t)$. Restricting to small $\Omega $ and $\dot{\Omega} $ meaning that the particle completes the circuit many times in time $T$ during which the loop revolves only once, the $s$-dependence can be averaged out to give  
\begin{equation}
\label{bb20}
s(T)=s_0+p_0T +\int_0^Tdt'~[\frac{1}{L}\int_0^L ds  (t-t')F(t',s)].
\end{equation}
The $s$ or equivalently angle averaging corresponds to the adiabetic principle \cite{arnold}. With proper scaling the last term in RHS of (\ref{bb20})  reproduces
 the anholonomy or Hannay angle. Below we will exploit both (\ref{h2}) and (\ref{bb20}) to compute  Hannay angles for different noncontractible loops on torus.

{\bf{(2.1) Toroidal loops:}} For toroidal loops (see Figure 1)    $\theta =\theta_0 $ is fixed,
\begin{equation}
\label{a1}
{\bf{r}}\equiv \{(c+a~cos\theta_0)cos \phi, ~(c+a~cos\theta_0)sin \phi,~a~sin\theta_0\};~~ {\bf{\Omega}}\equiv \{\Omega_1, ~\Omega_2,~\Omega_3\}.
\end{equation}	
We find
\begin{equation}
\label{h5}
\oint ({\bf \Omega}\times {\bf r})\bigcdot\,d{\bf r}=\int _0^{2\pi }d\phi  [-\Omega_2a~sin\theta_0 sin\phi -\Omega_1 a~ sin\theta_0 cos\phi +\Omega_3 (c+a~cos\theta_0)]$$$$=2\pi (c+a~cos\theta_0) ^2\Omega_3 \equiv 2A\Omega_3
\end{equation} 
where $A$ denotes area of the loop - a circle with radius $c+a~cos\theta_0$.
The last form is similar to the structure in \cite{h1} for a circular hoop since qualitatively toroidal loops with fixed $\theta_0$ is same as a circular hoop. Due to anholonomy, the (spatial) average speed is greater by $2A\Omega_3/L$ and for small $\Omega$ one equates spatial average to time average. Hence over a single revolution of the loop in $\Omega_3$, $\int_0^T \Omega_3dt=2\pi$ and the extra distance traversed is $-4\pi A/L$. Equivalently  the extra angle (Hannay angle)  is obtained by scaling by $2\pi/L$ to get 
\begin{equation}
\label{hh12}
\Delta \Theta =-\frac{8\pi^2 A}{L^2}.
\end{equation}
For circular loop this is simply $-2\pi$. The fact the Hannay angle depends only on $\Omega_3$ agrees with the general statement of Hannay \cite{h2} since the toroidal loop  in horizontal plane has zero projection on planes perpendicular to $\Omega_1$ and $\Omega_2$.

Now from Berry's method,  for (\ref{a1}) with 
${\bf t}=\{-sin\phi,~cos\phi ,~0\}$
we get from (\ref{h3})
\begin{equation}
\label{b1}
\ddot {s}=\frac{1}{2}(c+acos\theta_0) (\Omega_2^2-\Omega_1^2)sin(2\phi) +\Omega_1\Omega_2 (c+acos\theta_0) cos(2\theta )+a\Omega_3sin\theta_0(\Omega_2cos\phi -\Omega_1 sin\phi )$$$$+asin\theta_0(\dot {\Omega_2}sin\phi +\dot {\Omega_1}cos\phi)-(c+acos\theta_0) \dot{\Omega_3} .
\end{equation}
Only the last term with $\dot{\Omega_3}$ will survive the loop averaging to yield 
$$\Delta s= -\int_0^T dt' (T-t')(c+acos\theta_0) \dot{\Omega_3}. $$   
Using $\int_0^T dt' (T-t')\dot{\Omega_3}=2\pi$ we recover
\begin{equation}
\label{bb2}
\Delta s  =-2\pi (c+a~cos\theta_0).
\end{equation}
Now $\Delta\Theta=(2\pi/L)\Delta s=-2\pi $.
Note that this result is a straightforward extension of the hoop result since the closed path considered here for fixed $\theta=\theta_0$ is similar to the motion along a hoop.

{\bf{(2.2) Poloidal loops:}} For poloidal loops   $\phi=\phi_0$  fixed leading to
\begin{equation}
\label{h6}
\oint ({\bf \Omega}\times {\bf r})\bigcdot\, d{\bf r}=\int_0^{2\pi } d\theta [-\Omega _2(a~cos\phi_0 -c~cos\theta cos\phi_0)+\Omega_1(a~ sin\phi_0 +c~ cos\theta sin\phi_0)]$$$$=2\pi a^2(\Omega_1sin\phi_0-\Omega_2cos\phi_0) \equiv 2\bar A(\Omega_1sin\phi_0-\Omega_2cos\phi_0).
\end{equation}
 Here also the general statement of Hannay \cite{h2} is satisfied since  the poloidal loop in vertical  plane has vanishing projection on a plane perpendicular to $\Omega_3$ and Hannay angle depends only on $\Omega_1$ and $\Omega_2$.  The axial symmetry is absent which generates a  $\phi_0$-dependent result. For only non-zero $\Omega_1$ the anholonomy will vanish for $\phi_0=0,\pi $ because these two loops are on the axis of rotation of the loop and so do not experience the rotation effect. If either $\Omega_1$ or $\Omega_2$ is non-vanishing the Hannay angle follows similar arguments. However, if both $\Omega_1$ and $\Omega_2$ are non-vanishing the above argument is extended  to $\int_0^{\tilde T} \Omega_1dt=2\pi n_1,~\int_0^{\tilde T} \Omega_2dt=2\pi n_2$  with $n_1,n_2$ being smallest integers that satisfy the  condition that the torus returns to itself after revolution. Hence the Hannay angle is
\begin{equation}
\label{hh1}
\Delta \Theta =-\frac{8\pi^2\bar A}{L^2} (n_1 sin\phi_0-n_2 cos\phi_0).
\end{equation}
Notice that Hannay angle disappears for $n_1 sin\phi_0=n_2 cos\phi_0$.
Hannay angle for poloidal cycle, an example of noncontractible loop of a different type, is a new  and one of our major results.

We can deduce the same result utilizing  Berry's arguments where ${\bf t}\equiv \{-cos\phi_0 sin\theta ,~sin\phi_0 sin\theta, ~cos\theta \}$. We get
\begin{equation}
\label{b2}
\ddot{s}=-(\Omega_1^2cos^2\phi_0 +\Omega_2^2sin^2\phi_0+\Omega_1\Omega_2sin(2\phi_0)) (c+a~cos\theta )sin\theta $$$$+\frac{1}{2}a\Omega_3^2 sin(2\theta)+(\Omega_1 cos \phi_0+\Omega_2 sin \phi_0) \Omega_3(c~cos\theta+a~cos(2\theta ))-c~{\bf{\Omega}}^2sin\theta$$$$ -(\dot{\Omega_1} sin \phi_0-\dot{\Omega_2} cos \phi_0)(a+c~cos\theta ).
\end{equation}
Only the $a$-term in the last expression survives the $\theta$-integration and same procedure as discussed in the toroidal case will yield (\ref{hh1}).\\
{\bf{(2.3) $p,q$ Torus knot:}}
 The angle variables are parameterized  by $\phi \rightarrow p\phi,~\theta \rightarrow q\phi $ with $2\pi  \geq\phi \geq 0$, with $\omega = q/p$. From the position  vector,	
\begin{equation}
\label{k1}
{\bf{r}}\equiv \{(c+acos q\phi)cos p\phi; ~~(c+acos q\phi)sin p\phi;~~x_3=asin q\phi \}
\end{equation}	
we compute
\begin{equation}
\label{k2}
\oint{\bf{\Omega}}\times {\bf{r}}\bigcdot\, d{\bf{r}} =\int_0^{2\pi}d\phi~ p\Omega_3(c+acosq\phi)^2=p(2c^2+a^2)\pi \Omega_3.
\end{equation}
Total length of the $(p,q)$-knot is 
\begin{equation}
\label{kk2}
L=\int _ 0^ {2\pi} d\phi =[(\frac{dx_1}{d\phi})^2+(\frac{dx_2}{d\phi})^2+(\frac{dx_3}{d\phi})^2]^{1/2}=a\int_0^{2\pi}d\phi [q^2+p^2(n+cosq\phi)^2]^{1/2}
\end{equation}
and an approximate form can be obtained from the average of the maximum and minimum value of the expression:
\begin{equation}
L\approx\pi a ([q^2+p^2(n+1)^2]^{1/2}+[q^2+p^2(n-1)^2]^{1/2})
\label{k22}
\end{equation}
Hence a form of the Hannay angle is,
\begin{equation}
\label{hhh1}
\Delta \Theta =-\frac{2\pi^2}{L^2}p(2c^2+a^2)\pi .
\end{equation}
 The involved nature of the unit tangent vector makes the computation  more involved in Berry's framework, 
\begin{equation}
\label{k4}
{\bf{t}}\equiv \frac{1}{[q^2a^2+p^2(c+acos q\phi )^2]^{1/2}]}$$$$\{(-aq~sinq\phi~cosp\phi -p((c+acos q\phi )sinp\phi),~(-q~sinq\phi~sin p\phi +p((c+acos q\phi )cosp\phi),~q~acos q\phi\}.
\end{equation}
For the special case  ${\bf{\Omega}}=\Omega_3\hat k$, (since we have seen from (\ref{k2}) that only $\Omega_3$ contributes),
\begin{equation}
\label{k5}
\ddot s=-\frac{p(c+acos q\phi )^2}{[q^2a^2+p^2(c+acos q\phi )^2]^{1/2}}\dot{\Omega_3}-\frac{(c+acosq\phi )aqsinq\phi }{{[q^2a^2+p^2(c+acos q\phi )^2]^{1/2}}}\Omega_3^2
\end{equation}
and it is clear that only the first term ($\sim \dot{\Omega_3}$) will survive the loop integration. Comparing with (\ref{k2}) the structures are obviously similar but we do not pursue with this computation any further and remain satisfied with (\ref{hh1}) as the Hanny angle for the $(p,q)$-knot.
\vskip .3cm
{\bf{(3) Berry phase for noncontractible loops}}\\
From the knowledge of the quantum eigenstates of a system, the adiabatic phase $\gamma_n$ accumulated (for the $n$'th eigenstate $\psi_n({\bf{r}},{\bf{X}})$) after one rotation of the system parameters ${\bf{X}}$ is given by \cite{b3},
\begin{equation}
\label{bh1}
\gamma_n=-Im\int_{0}^{2\pi}d{\bf{X}}\int_{-\infty}^{+\infty}dx\int_{-\infty}^{+\infty}dy~\psi_n^*({\bf{r}},{\bf{X}})\frac{\partial}{\partial {\bf{X}}} \psi_n({\bf{r}},{\bf{X}}).
\end{equation}
The connection between Hannay angle $\Delta\Theta$ and Berry phase $\gamma_n$ was revealed in \cite{b3}:
\begin{equation}
\label{bhb1}
\Delta\Theta =-\hbar\frac{\partial}{\partial n}\gamma_n .
\end{equation}
Since non-zero Hannay angle guarantees a non-zero Berry phase this motivates us to look for the latter in the present context.

The Hamiltonian for a particle in a frame  rotating with ${\bf{\Omega}}$ relative to an inertial frame can be expressed as \cite{anan} $H=\frac{1}{2}({\bf{p}}-{\bf{\Omega}}\times {\bf{r}})^2-\frac{1}{2}({\bf{\Omega}}\times {\bf{r}})^2$, with an  effective gauge field identified as  $A^\mu =\{-\frac{1}{2}({\bf{\Omega}}\times {\bf{r}})^2,{\bf{\Omega}}\times {\bf{r}}\}$. However, in the present work, we have devised a new scheme   to obtain the wavefunctions in a rotating frame approximately  where we solve the Schrodinger equation in a static frame and then modify the variable to incorporate the rotating frame effect. 
\vskip .3cm
{\bfseries{(3.1) Toroidal loop:}} For toroidal loop $H$ reduces to
\begin{equation}
\label{bf3}
H=\frac{1}{2}\frac{p_\phi^2}{(c+acos\theta_0 )^2})=\frac{1}{2}\bar{p_\phi}^2
\end{equation}
where $\{p_{\bar\phi },\bar \phi =(c+acos\theta_0 )\phi \}$ constitute the canonical pair. For this  rotor (on a static torus) the eigenfunctions are 
\begin{equation}
\label{b3}
\psi_n({\bf{r}}, \bar\phi)=(a({\bf{r}}, \bar\phi)/\sqrt{L})exp[2\pi in\bar{\phi}/L]=(a({\bf{r}}, \bar\phi)/\sqrt{L})exp[2\pi in (c+acos\theta_0 )\phi/L],
\end{equation}
where $a^2$ has a $\delta$-function like behavior being non-zero only on the path \cite{b3}.  Now we modify the static rotor wavefunction since the torus itself is rotating with ${\bf{\Omega}}$. Let us first introduce the arc length $s$ by using  $\phi =2\pi s/L$,
\begin{equation}
\label{bh3}
\psi_n({\bf{r}}, \bar\phi)=(a({\bf{r}}, \bar\phi)/\sqrt{L})exp[4\pi^2 in (c+acos\theta_0 )s/L^2].
\end{equation}
Direction of the particle motion in toroidal loop is given by,
${\bf{s}}=s\hat{{\bf{\phi}}}=s\frac{\partial {\bf{r}}}{\partial \phi}/\mid \frac{\partial {\bf{r}}}{\partial \phi}\mid $. Following our earlier analysis in Section 2 \cite{h1}
\begin{equation}
	\label{bp1}
	\frac{d{\bf{s}}}{dt}\rightarrow  ({\bf{\Omega}}\times  {\bf{s}}) \bigcdot\, \hat{\bf{r}}=s ({\bf{\Omega}}\times  \hat{\bf{\phi}})\bigcdot\, \hat{\bf{r}}= -\frac{s(c+acos\theta_0)\Omega_3}{\sqrt{c^2+a^2+2cacos\theta_0}}.
\end{equation}
The Berry phase is computed as
\begin{equation}
\label{b4}
\gamma _n=-Im\int_0^{L}ds\int_{-\infty}^{\infty}dx\int_{-\infty}^{\infty}dy~\psi_n^*\frac{\partial}
{\partial s}\psi_n =[\frac{4\pi n A }{L^2}][\frac{L}{\sqrt{c^2+a^2+2cacos\theta_0}}]$$$$ \approx [\frac{8\pi^2n A }{L^2}][1-\frac{1}{2}\frac{a^2sin^2\theta_0}{(c+acos\theta_0)^2}],
\end{equation}
where $A=\pi (c+acos\theta_0)^2$ and  we have taken time averaging for one complete revolution of the torus $\frac{1}{2\pi}\int_0^Tdt\Omega_3=1 $. The leading term  agrees with the Hannay angle  (\ref{hh12}) in the thin torus limit $c>>a$.
\vskip .3cm

{\bfseries{(3.2) Poloidal loop:}} Coming to the poloidal loop $H$ simplifies  to
\begin{equation}
\label{b5}
H=\frac{1}{2a^2}p_\theta^2=\frac{1}{2}\bar{p_\theta}^2
\end{equation}
with $\{p_{\bar\theta},\bar \theta =a\theta \}$ being the canonical pair. Let us write the rotor $\psi_n({\bf{r}},{\bf{X}})$ as
\begin{equation}
\label{bh3}
\psi_n({\bf{r}}, \bar\phi)=(a({\bf{r}}, \bar\phi)/\sqrt{L})exp[2\pi in\bar{\theta}/L]=(a({\bf{r}}, \bar\phi)/\sqrt{L})exp[2\pi in a\theta /L ]
\end{equation}
For poloidal path the particle moves along ${\bf{s}}=s\hat{{\bf{\theta}}}=s\frac{\partial {\bf{r}}}{\partial \theta}/\mid \frac{\partial {\bf{r}}}{\partial \theta}\mid $. 
Once again the torus revolution factor requires that  $s$  be replaced by 
\begin{equation}
\label{bph1}
\frac{d{\bf{s}}}{dt}\rightarrow  ({\bf{\Omega}}\times  {\bf{s}})\bigcdot\, \hat{\bf{r}}=s ({\bf{\Omega}}\times  \hat{\bf{\theta}})\bigcdot\, \hat{\bf{r}}= -\frac{s(a+ccos\theta )(\Omega_2 cos\phi_0 - \Omega_1 sin\phi_0)}{\sqrt{c^2+a^2+2cacos\theta}}.
\end{equation}
It is interesting to note that the characteristic factor $\Omega_2 cos\phi_0 - \Omega_1 sin\phi_0$ encountered earlier in (11) has appeared. However notice that due to the $\theta$-dependence in (\ref{bph1}) the angular integral is nontrivial giving the factor $2\pi c/a$ in the thin torus approximation. The Berry phase is obtained as
\begin{equation}
\label{bbh1}
\gamma_n=-\frac{4\pi n\bar A}{L^2} (n_1 sin\phi_0-n_2 cos\phi_0)[2\pi\frac{c}{a} +...].
\end{equation}
It is intriguing to note that, due to the non-trivial $\theta $-integral, there is a  mismatch between (\ref{hh1}) and (\ref{bbh1}) by the factor $c/a$,   even in the thin torus limit. We will comment on this point in Section 4 at the end.
\vskip .3cm

{\bf{(3.3) $p,q$ Torus knot:}} The situation is much more complicated for the knot with
\begin{equation}
\label{b7}
H=\frac{p_{\phi}^2}{2f(\phi )},~~f(\phi )=a^2\omega^2+(c+acos\omega\phi )^2,
\end{equation}
since the scaling needed for $p_\phi $ to bring $H$ to a canonical form is $\phi$ dependent and  not numerical, (as was the cases for toroidal and poloidal loops). On the other hand following \cite{sree} it is straightforward to compute the eigenvalues and eigenfunctions, at least for cases with $c>>a$. The solution of the Schrodinger equation for the wave function $\psi =\sqrt{f}\Sigma$ can be obtained from 
\begin{equation}
\label{s1}
(\frac{d^2}{d\phi ^2}+U(\phi ))\Sigma =0
\end{equation}
where 
\begin{equation}
\label{b8}
U(\phi )=\frac{2ff''-(f')^2}{12f^2}+\frac{2Ef}{\hbar^2},~~f'=df/d\phi .
\end{equation} 
 Approximating for a thin torus, $c>>a$, we find
\begin{equation}
\label{s2}
U(\phi )\approx \frac{\lambda \omega ^2}{4}-\frac{\omega^2}{3\sigma }cos(\omega \phi)
\end{equation}
where $\lambda =\frac{8c^2E}{\hbar^2\omega^2},~a/c=1/\sigma $. The equation is in the form of Mathieu equation. The energy eigenvalues and eigenfunctions, to $O(1/\sigma )$ are 
\begin{equation}
\label{s3}
E_n=\frac{n^2\hbar^2\omega^2}{2c^2q^2}$$$$
\psi_{+}^{(n)}\approx (c/\sigma) cos(n\omega\phi /q)+O(1/\sigma^2)=acos(n\phi/p)+O(1/\sigma^2),$$$$
\psi_{-}^{(n)}\approx (c/\sigma) sin(n\omega\phi /q)+O(1/\sigma^2)=acos(n\phi/p)+O(1/\sigma^2)
\end{equation}
with periodicity $2\pi p$ in $\phi $. Hence, following \cite{b3} we consider wavefunctions  with proper periodicity, of the form
\begin{equation}
\label{s4}
\psi_n =(A/\sqrt L)exp[2\pi i n s/L].
\end{equation}

To  account for the revolution of the torus itself  we go back to the knot parameterization (\ref{k1}) 
with  particle motion along
${\bf{s}}=s\hat{{\bf{\phi}}}=s\frac{\partial {\bf{r}}}{\partial \phi}/\mid \frac{\partial {\bf{r}}}{\partial \phi}\mid $. (Note that this $\hat{\phi}$ has to be derived from (\ref{k1}) above and should not be confused with the $\hat{\phi}$ used earlier for toroidal loop.) Following our earlier analysis in Section 3 \cite{h1}
\begin{equation}
\label{bbp1}
\frac{d{\bf{s}}}{dt}\rightarrow  ({\bf{\Omega}}\times  {\bf{s}})\bigcdot\, {\bf{r}}/(\mid  {\bf{s}}\mid \mid  {\bf{r}}\mid ).
\end{equation}
We explicitly compute 
\begin{equation}
\label{00}
({\bf{\Omega}}\times  {\bf{s}})\bigcdot\, {\bf{r}}=a(c+acosq\phi )[\Omega_1(p~sinq\phi~ cosp\phi-q~sinp\phi~ cosq\phi)+\Omega_2(p~sinq\phi~ sinp\phi+q~cosp\phi~ cosq\phi)]$$$$+a^2qsin^2q\phi [-\Omega_1 sinp\phi +\Omega_2 cosp\phi ]+\Omega_3p(c+acosq\phi )^2,
\end{equation}

\begin{equation}
\label{kn2}
\mid  {\bf{s}}\mid \mid  {\bf{r}}\mid =[a^2q^2+(c+acosq\phi )^2 p^2]^{1/2}[a^2+c^2+2accosq\phi ]^{1/2} .
\end{equation}
Hence for $c>>a$ (thin torus approximation) in the leading order the only surviving term in (\ref{bbp1}) is the last term in rhs of (61) that is $\Omega_3$. The leading order result agrees with our previous result in (\ref{k2}) where also only $\Omega_3$ contributed. We again exploit  time averaging for one complete revolution of the torus to remove $\Omega_3$, $\frac{1}{2\pi}\int_0^Tdt\Omega_3=1 $. After this intermediate step, from (\ref{s4}),  
 The Berry phase turns out to be
\begin{equation}
\label{sbh1}
\gamma_n=-Im\int_{0}^{2\pi p}d{\bf{X}}\int_{-\infty}^{+\infty}dx\int_{-\infty}^{+\infty}dy~\psi_n^*({\bf{r}},{\bf{X}})\frac{\partial}{\partial {\bf{X}}} \psi_n({\bf{r}},{\bf{X}})$$$$
=-\frac{2\pi n}{L^2}\int^{2\pi p}_0 d{\bf{X}}\int_0^L ds \frac{\partial s}{\partial X}\approx -\frac{2\pi n}{(2\pi p c)^2}(2\pi p) \pi c^2.
\end{equation}
Indeed, the above is a crude approximation where we have considered the area enclosed by the knot simply as a disc of radius $c$ (since $c>>a$) but take length of loop as  $L\approx 2\pi c p$, the $c>>a $ limit of (\ref{k22}) as the particle traversed the $c$-loop $p$ times. In this limit the motion in the poloidal direction becomes insignificant. This agrees with our previous result for Hannay angle (\ref{hhh1}) in the limit $c>>a $.

\vskip .3cm
{\bf{(4) Conclusion and future directions}}\\
To summarize we have computed  Hannay angle and Berry phase for noncontractible  loops on a slowly revolving  torus The closed trajectories are in the form of toroidal and poloidal cycles and knots (containing both these forms of loops). These type of loops have not been studied before in the present context. 

We have explicitly verified a general observation by Hannay concerning the relation between Hannay angle, the loop itself and the rotation axis of the loop. The Hanny angles for particle moving on  noncontractible cycles when the torus itself is revolving are computed. We have used two distinct frameworks suggested by Hannay \cite{h1} and Berry \cite{b2} and the results agree (albeit for knots where we have indicated the agreement in leading order). A new and interesting result is that there is more structure in the Hannay angle expression for poloidal loop, not anticipated earlier, although the toroidal loop result is  essentially similar to Hannay's result for a planar hoop \cite{h1}. 

Next we have moved over to the quantum regime and calculated Berry phase corresponding to the above paths. We have compared the Hanny angle and Berry phase by exploiting the relation proved by Berry \cite{b2}. For toroidal loops (and for knots at a limit $c>>a$, $c$ and $a$ being the radii of the toroidal circle and poloidal circle respectively) the results agree. However, in case of the poloidal loop we observe that the Berry phase and Hannay angle differ by a factor $\sim c/a$ which needs to be commented upon. We believe that the classical result (\ref{hh1}) is correct and the fault presumably lies with the result  (\ref{bbh1}) computed in a semi-classical framework. A clear reason for this mismatch is that, of all the non-contractible loops on a torus, all poloidal loops are topologically equivalent (they can be simply slided along the inner  surface to map one onto the other). Same is true for all   toroidal loops that can be slided on the  outer surface. However, poloidal and toroidal loops are topologically distinct since  a poloidal loop can not be mapped onto a toroidal loop without cutting open the torus. Obviously, in a qualitative way the toroidal loops are similar to the easier problem of particle on a (irregular) hoop and so in these cases the semi-classical analysis suffices. On the other hand as we have demonstrated the semi-classical framework is not sufficient for poloidal loops. Indeed, this is not totally surprising but for the first time in the present work this difference has been exposed in an explicit way. The need for a full quantum mechanical treatment, possibly utilizing the quantum potential generated by the constrained dynamics of the particle (fixed path in configuration space) becomes imperative. It appears that the  general framework of such computation, as provided in \cite{QM}, will be relevant in the present context. Since this, in itself, is an independent problem we have not included it in the present work and it will be communicated in a future publication.

Let us list some of the interesting  future directions of work.\\
(i) We are pursuing the quantum computation of the Berry phase for particle moving on a poloidal loop taking in to account constrained motion of the particle in to account.\\
(ii) It will be very interesting to introduce a magnetic field and study the revolving torus along with the magnetic field's effect on Berry phase and Hannay angle. The magnetic field can be of poloidal, toroidal or arbitrary nature. \\
(iii)  A natural and non-trivial generalization would be to consider a generic form of torus that is, a deformed torus, compared to the canonical (regular) form of torus studied here. This will indicate how general (or topological) our results are. A possible way to do it is to work with a toroidal coordinate system and then try to see effects of small variations about fixed $c$ and $a$. The same goes for generic nature of loops as well. Another interesting aspect is to generalize the present analysis to higher genus surfaces.\\
(iv) A non-trivial demonstration of the topology of the two-torus, (on which the particle trajectories are embedded), would be to consider particle paths that cover the whole toroidal surface with the torus itself subjected to adiabatic rotation about the symmetry axis. The dynamics will still be integrable. In this case there will appear two Hannay angles (instead of one for paths on a  sphere). \\
(v) As a practical realization in the classical context one can think about a particle possessing a constant vector fixed on it such as a spinning top or a gyroscope, (though in case of top the spin vector is radial whereas we  have considered the vector to have vanishing radial component), with the particle sliding over a torus. In an optical setup one can think of light moving in a  torus knot loop of  optical fibre \cite{h2}. It is an open problem for experimentalists to construct time dependent magnetic fields that can simulate noncontractible cycles in parameter space.\\

\vspace{.1cm}
{\bf{Acknowledgments:}} It is a pleasure to thank Professor John Hannay for many helpful correspondences and concrete suggestions. Also I am grateful to Professor Michael Berry for comments. Also I thank the referee for the constructive comments and suggestions.

\vskip 1cm
Ethics statement:  This work did not involve any active collection of human data.\\
Data accessibility statement: This work does not have any experimental data.\\
Competing interests statement: We have no competing interests.\\
Authors’ contributions: There is only one author. \\
Funding: This work was not supported by any funding agency.\\


\begin{thebibliography}{99}
\bibitem{monod} 	The situations where   where noncontractibility plays an essential role arise when the curvature vanishes and yet there are still nontrivial anholonomies. In this case,
	 they are called monodromies, meaning the absence of a smoothly varying set of quantum numbers that characterize the system.   The monodromy of a closed loop depends 	 only on its homotopy class and constitute a representation 	 (possibly nonabelian) of the fundamental group. 	    
		 In an early work Robbins and Berry \cite{b2} have provided an example of Berry phase for spin system in   magnetic field that reverses in direction and thus completes half a cycle without enclosing any area. It was shown that for $m=0$ spin state the Berry phase vanished for trivial (contractible) cycles whereas it is non-zero for non-trivial cycles. I thank Professor Berry for informing me about this reference.
\end{thebibliography}
\end{document}